\documentclass[conference]{IEEEtran}
\IEEEoverridecommandlockouts
\usepackage{cite}
\usepackage{amsmath,amssymb,amsfonts}
\usepackage{algorithmic}
\usepackage{graphicx}
\usepackage{textcomp}
\usepackage{xcolor}
\usepackage{balance}
\def\BibTeX{{\rm B\kern-.05em{\sc i\kern-.025em b}\kern-.08em
    T\kern-.1667em\lower.7ex\hbox{E}\kern-.125emX}}
\begin{document}

\title{M2M-Gen: A Multimodal Framework for Automated Background Music Generation in Japanese Manga Using Large Language Models
}


\author{\IEEEauthorblockN{Megha Sharma}
\IEEEauthorblockA{\textit{Department of ICE} \\
\textit{The University of Tokyo}\\
Tokyo, Japan \\
meghas@g.ecc.u-tokyo.ac.jp,\\
ms22sharma@gmail.com}
\and
\IEEEauthorblockN{Muhammad Taimoor Haseeb}
\IEEEauthorblockA{\textit{Music X Lab} \\
\textit{MBZUAI}\\
Abu Dhabi, UAE \\
taimoor.haseeb@mail.mcgill.ca}
\and
\IEEEauthorblockN{Gus Xia\textsuperscript{*}\thanks{\textsuperscript{}* The last two authors are senior authors.}}
\IEEEauthorblockA{\textit{Music X Lab} \\
\textit{MBZUAI}\\
Abu Dhabi, UAE \\
gus.xia@mbzuai.ac.ae}
\and
\IEEEauthorblockN{Yoshimasa Tsuruoka\textsuperscript{*}}
\IEEEauthorblockA{\textit{Department of ICE} \\
\textit{The University of Tokyo}\\
Tokyo, Japan \\
yoshimasa-tsuruoka@g.ecc.u-tokyo.ac.jp}
}

\maketitle 

\begin{abstract}
This paper introduces M2M-Gen (Manga-to-Music Generation), a multi-modal framework for generating background music tailored to Japanese manga. The key challenges in this task are the lack of an available dataset or a baseline. To address these challenges, we propose an automated music generation pipeline that produces background music for an input manga book. Initially, we use the dialogues in a manga to detect scene boundaries and perform emotion classification using the characters’ faces within a scene. Then, we use GPT-4o to translate this low-level scene information into a high-level music directive. Conditioned on the scene information and the music directive, another instance of GPT-4o generates page-level music captions to guide a text-to-music model. This produces music that is aligned with the manga’s evolving narrative. The effectiveness of M2M-Gen is confirmed through extensive subjective evaluations, showcasing its capability to generate higher quality, more relevant and consistent music that complements specific scenes when compared to our baselines. We provide output samples of the pipeline at: manga-to-music.github.io/M2M-Gen/

\end{abstract}


\section{Introduction}
Music stimulates experiences, creating richer and more dynamic narratives by integrating sound with visual storytelling, as exemplified by the soundtracks in anime and video games. In fact, the concept of music accompanying comics is not novel; Blin’s \cite{blin2019becoming} analysis of the French comic art form \textit{bande dessinée} demonstrates that music has long been integral to the comic creation process. From a reader's perspective, music and manga (Japanese comics) are often shared interests \cite{stevens2008japanese}, underscoring the importance of exploring the relationship between comics and music. Online comic platforms such as Webtoon\footnote{https://www.webtoons.com/en/} use this to their advantage by increasingly incorporating background music scores and sound effects as users scroll through the comics.

Creating background music for comics demands a novel composition that is relevant to the changing mood, theme, and pace, requiring composers to be highly creative and adaptable while collaborating with comic artists. For many artists, this memorable addition is exclusive to those who can find and employ musicians. In the case of comic background music, little to no literature supports the exploration of music generation using machine learning, which leaves the field with an open horizon of opportunities. This is partly because creating fitting music pieces for comic scenes that evolve with the unfolding story is a complex and challenging problem. In this work, we introduce a novel task of background music generation for manga. Lacking an end-to-end dataset and prior research on this specific task, we explore this domain and establish a baseline to encourage future research and exploration.

Leveraging recent advancements in vision, text, and music generation models, we formalize and explore the relationship between manga and music. The proposed system, M2M-Gen (Manga-to-Music Generation), is designed to generate background music that aligns with the content and sentiment of the manga panels, evolving as the narrative unfolds. This is particularly useful for manga artists and anime music directors to automate the music generation process. While the system is fully autonomous, the pipeline is designed to allow for human intervention at any step. This is mindful of human creativity and respects it as the highest caliber of the music generation process. Moreover, our step-by-step pipeline structure establishes that the text output from each step ensures transparency, is human-readable, and can trace errors in the system. Our key contributions are summarised below:

\begin{itemize}
    \item We introduce a novel task for generating background music for manga and present a pipeline that successfully implements this task.
    \item We evaluate our work through extensive user studies, comparing our pipeline against a strong baseline and a random lower bound.
    \item In our modular approach, all but one model is publicly available, making our pipeline efficient for generating music for low-resource comics with minimal training.
\end{itemize}

\begin{figure*}[ht!]
\centering
    \includegraphics[width=1.0\textwidth]{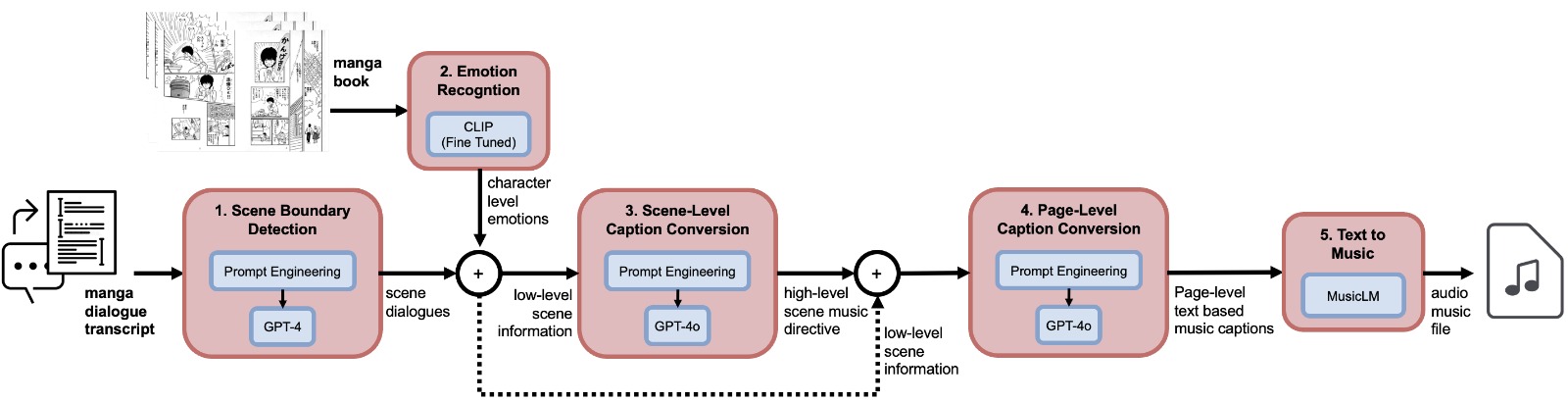}
    \caption{Pipeline for M2M-Gen. It takes as input the images and dialogue transcript of a manga book and outputs an audio file containing tailored background music. Manga image taken from Manga109 
    \cite{mtap_matsui_2017} \cite{multimedia_aizawa_2020} 
    courtesy of Yoshi Masako.}
    \label{fig:m2mgen1}
\end{figure*}

\section{Related Work}
Current literature lacks studies on the generation of background music for manga or other comics. Insights from research on music generation for games \cite{hutchings2019adaptive}, books \cite{shriram2022sonus}, and stories \cite{won2021emotion} indicate that emotional information can be crucial for generating a musical score for any medium. Apart from emotion, rich context can play a key role in generating background music for multi-modal media. Haseeb et al. \cite{haseeb2024gpt} introduced a movie-to-music pipeline named Hermann-1 that inputs a movie scene and, through contextual and emotional information, develops the text prompt for a text-to-music model. The generated music complements the movie scene and achieves comparable results to the original soundtrack of the movie scene. Manga, like movies, also engages with its audience using images and dialogue. In this work, we draw inspiration from Hermann-1, customizing its components to adapt to the structure of manga.

In another study exploring background music for manga, Sharma et al. \cite{sharma_2023_10114097} fine-tuned AudioCLIP \cite{guzhov2022audioclip} to infer the relationship between manga and music based on their genre and mood information, respectively. However, their proposed model assumes that the emotion of an entire page can be summarized in one word, which may not always be accurate. In fact, a richer context and deeper understanding of the sentiment in the manga are crucial to achieving relevance and consistency with the narrative. Moreover, using retrieval as a solution for background music is limited by the quality and scope of the available dataset, and by the consistency that can be achieved between independent pre-recorded music pieces.

To the best of our knowledge, we are the first to address the challenge of generating background music for manga.

\section{Methodology}
A pipeline based approach is reasonable for tasks lacking an end-to-end dataset to objectively measure performance. Our method leverages existing pre-trained models to better understand an input manga.  First, we divide the book into scenes that follow narrative continuity, identified by changes in time, location, and/or character. For each scene, we extract visual, emotional, and dialogue information, which is processed by a Large Language Model (LLM) to generate a high-level scene music directive. Finally, we create music captions for each page in the scene, which are used as input for a text-to-music model. In addition to comparing with randomly selected music, we also introduce a baseline to assess M2M-Gen.

We implement this methodology using the Manga109 dataset \cite{mtap_matsui_2017} \cite{multimedia_aizawa_2020}, which includes transcribed dialogues with speaker information \cite{li2023manga109dialog}. The M2M-Gen pipeline is shown in Fig. \ref{fig:m2mgen1}. The components are described below:

\textit{\textbf{1. Scene Boundary Detection.}} 
We conducted prompt engineering on GPT-4 to dissect an input manga book into scenes using the dialogue transcript. The model is conditioned with an example scene boundary of 90 lines of dialogue, from the Manga109 dataset, annotated internally in the absence of a ground truth. Given the dialogues of an input manga book, the model is prompted to predict the scene boundaries. 

\textit{\textbf{2. Emotion Recognition.}} As exemplified in our literature review, understanding the emotional tone of a scene can be crucial for creating appropriate background music. Our work explicitly detects and utilizes facial emotion recognition to guide the music generation process. To train on this task, we use KangaiSet \cite{theodose2023kangaiset}, a dataset containing manga characters' face and their emotion. We experimented with various vision models like ResNet \cite{he2016deep}, EfficientNet \cite{tan2019efficientnet}, and Vision Transformer \cite{dosovitskiy2020image}, achieving our best results with CLIP \cite{radford2021learning}. We use the Adam optimizer with betas set to 0.9 and 0.98 and a weight decay of 0.001. The learning rate is set to $1e-5$, and the model is trained for 10 epochs with a batch size of 16.

\textbf{\textit{3. Scene-Level Caption Conversion.}} This step leverages GPT-4o's ability to recognise musical directives given low-level scene information. Our system prompt consists of the following important information: 1) Instructions to create music directives for a manga scene, 2) Samples of music directives used in text-to-music models, and 3) The ideal length of the musical description. We adhere to a polite sentence structure, incorporating words like ``please” as recent research suggests that LLMs perform better with moderately polite prompts  \cite{yin2024should}. We optimized this approach through extensive prompt engineering. High-level scene music directive samples and input prompts are available on our demo page.

\begin{figure*}[ht!]
\centering
    \includegraphics[width=1.0\textwidth]{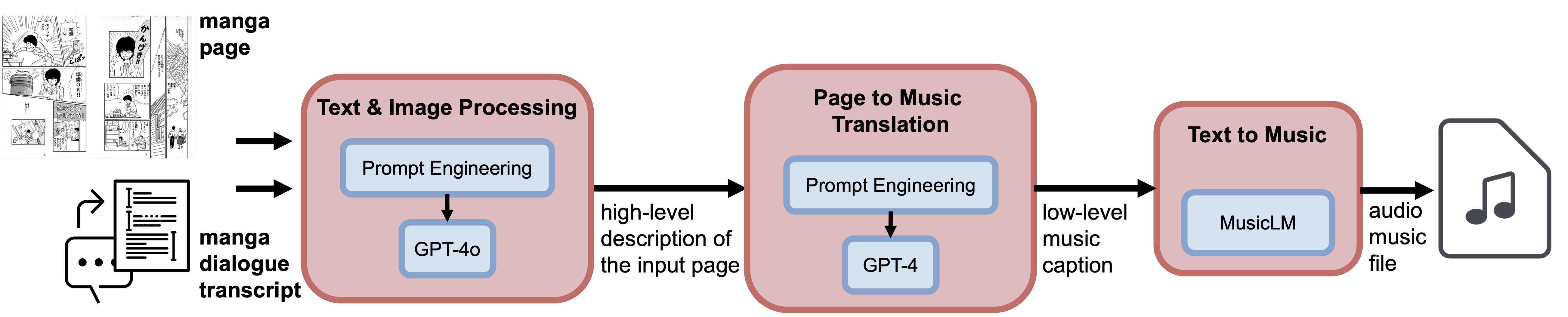}
    \caption{Baseline pipeline. It takes as input a manga page and dialogue transcripts and outputs an audio file containing tailored background music for the page. Manga image taken from Manga109 
    \cite{mtap_matsui_2017} \cite{multimedia_aizawa_2020} 
    courtesy of Yoshi Masako.}
    \label{fig:baseline}
\end{figure*}

\textbf{\textit{4. Page-Level Caption Conversion}}. To ensure that each page within a scene accurately reflects its unique emotional tone, we further break down the caption to page-level. This step converts a scene music directive into a page music caption using an instance of the GPT-4o model with its role set as a music composer. Our prompt consists of the outputs of the three previous steps: 1) Scene dialogues, 2) Emotion recognition results, and 3) Music directive of the scene. This approach guarantees that the generated music captions are diverse, consistent, and relevant across pages within the same scene, adapting to the evolving emotions. Sample low-level music captions are available on our demo page.

\textbf{\textit{5. Text-to-Music}}. In the final phase, we generate music for a page using page-level text-based music captions. We evaluated state-of-the-art text-to-music models, including MusicGen \cite{copet2024simple} and MusicLM \cite{agostinelli2023musiclm}. Ultimately, we chose MusicLM for our pipeline, as it received more favorable feedback. To ensure a seamless music transition between pages, we experiment with in-painting, conditioning on music generated for the previous page, and fading between musical pieces.

\section{Evaluation}

Evaluating M2M-Gen is challenging due to the lack of an end-to-end dataset or baseline for comparison. Using random music to evaluate M2M-Gen may be too simplistic. Therefore, we establish a baseline method with minimal engineering, illustrated in Fig. \ref{fig:baseline}. The baseline has three steps: 1) GPT-4o describes a Manga page using the image and the dialogue, 2) The high-level description is converted into a low-level music caption using another instance of GPT-4o, and 3) This musical description is fed into a text-to-music model. We fade the music between pages for a seamless transition.
 
Our final results are derived from a subjective analysis conducted through user studies. Since there is no dataset for this task, an objective assessment of the whole pipeline becomes infeasible. Finally, we report results of fine-tuning a component in the pipeline. 

\begin{figure}[ht]
\centering
    \includegraphics[width=1.0\columnwidth]{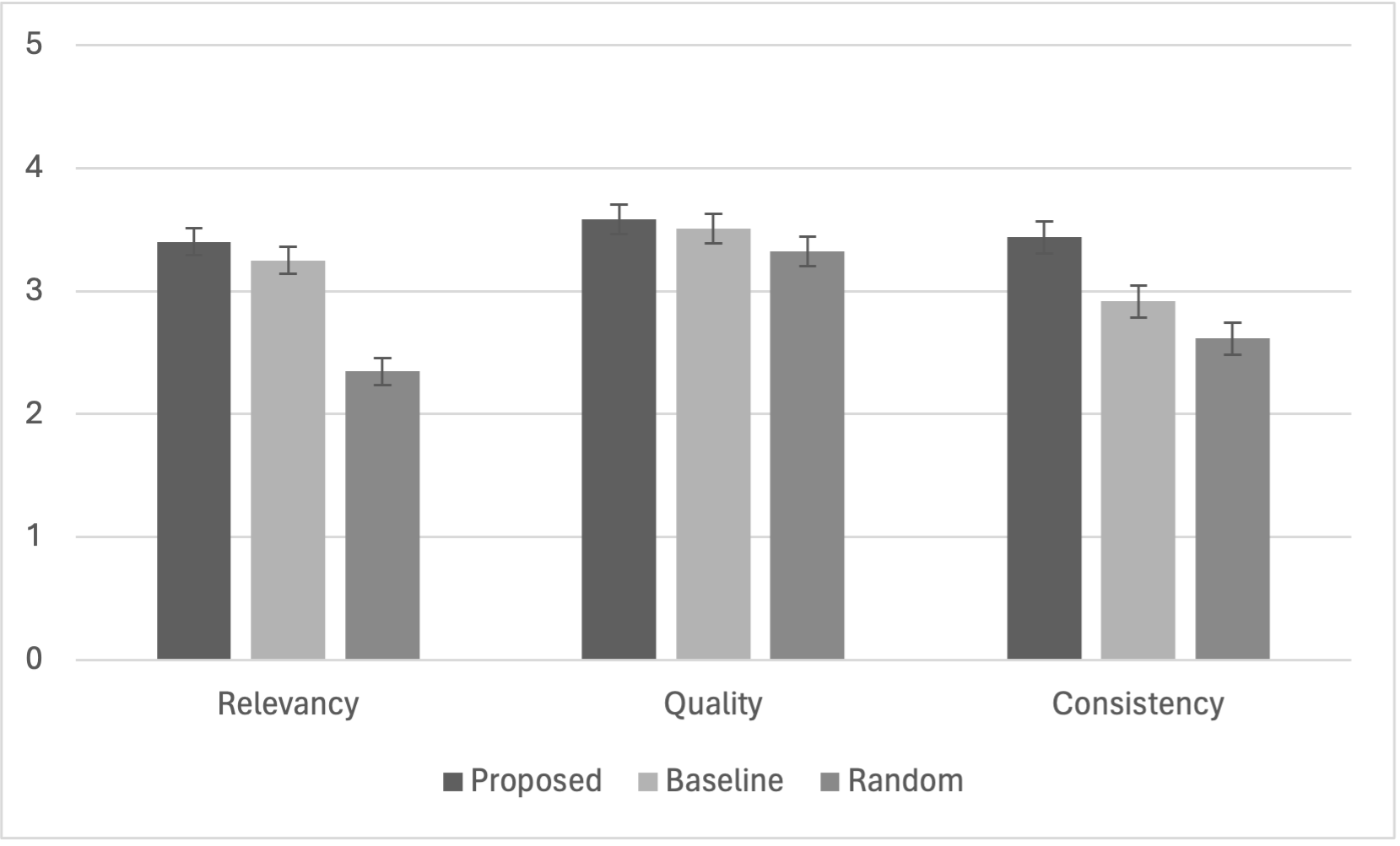}
    \caption{Final mean scores for each of the metrics using within-subject ANOVA on the within-scene study}
    \label{fig:finalresults}
\end{figure}

\subsection{Subjective Evaluation}
\textit{\textbf{Study Design.}} 
We follow a study design similar to Yang et al. \cite{yang2019deepmusicanalogylatent}. Each participant evaluated three different versions of six manga scenes, featuring background music from: 1) M2M-Gen, 2) baseline, and 3) a randomly selected track from a database of machine-generated music which acts as a lower bound. Each stimulus features four pages from a scene identified by GPT-4o in the scene boundary detection step. The survey involved 22 participants, all fluent in Japanese. All participants expressed their interest in manga and music to be at least above 2 (little interested) on a 5-point scale. All participants were recruited on a voluntary basis. To prevent bias, the scenes were shown in a random order, and participants were unaware of the music source. Each participant rated all eighteen samples on a 5-point Likert scale, from 1 (very low) to 5 (very high), based on the following criteria:

\begin{itemize}
    \item \textbf{Relevancy:} How appropriate is the music for the scene? 
    \item \textbf{Quality:} How ``good" does the music sound?
    \item \textbf{Consistency:} How well does the music flow?
\end{itemize}

\textit{\textbf{Results and Discussion.}} The  results are illustrated in Fig. \ref{fig:finalresults}. Both the baseline and proposed pipeline perform well above the random lower bound. Across all three questions, the proposed pipeline is preferred over the baseline and random method with statistical significance $(p < 0.001)$. The most notable difference is in relevance, where the proposed pipeline $(\mu = 3.4)$ and the baseline $(\mu = 3.25)$ outperform the random method $(\mu = 2.34)$ by nearly one point. In terms of quality, the differences are less pronounced but interesting nonetheless. Although the difference in mean quality scores is smaller, this difference is unexpected given that the music generation model remained constant across all methods. This implies that not only did participants find the generated outputs of M2M-Gen more relevant, but on average participants also perceived the quality of the proposed pipeline to be better than the rest. Perhaps the greatest contribution of M2M-Gen lies in the consistency of the generated output, with a half-point difference in average scores compared to the baseline. Here, the baseline and random methods are comparable.

\begin{figure}[ht]
\centering
    \includegraphics[width=1.0\columnwidth]{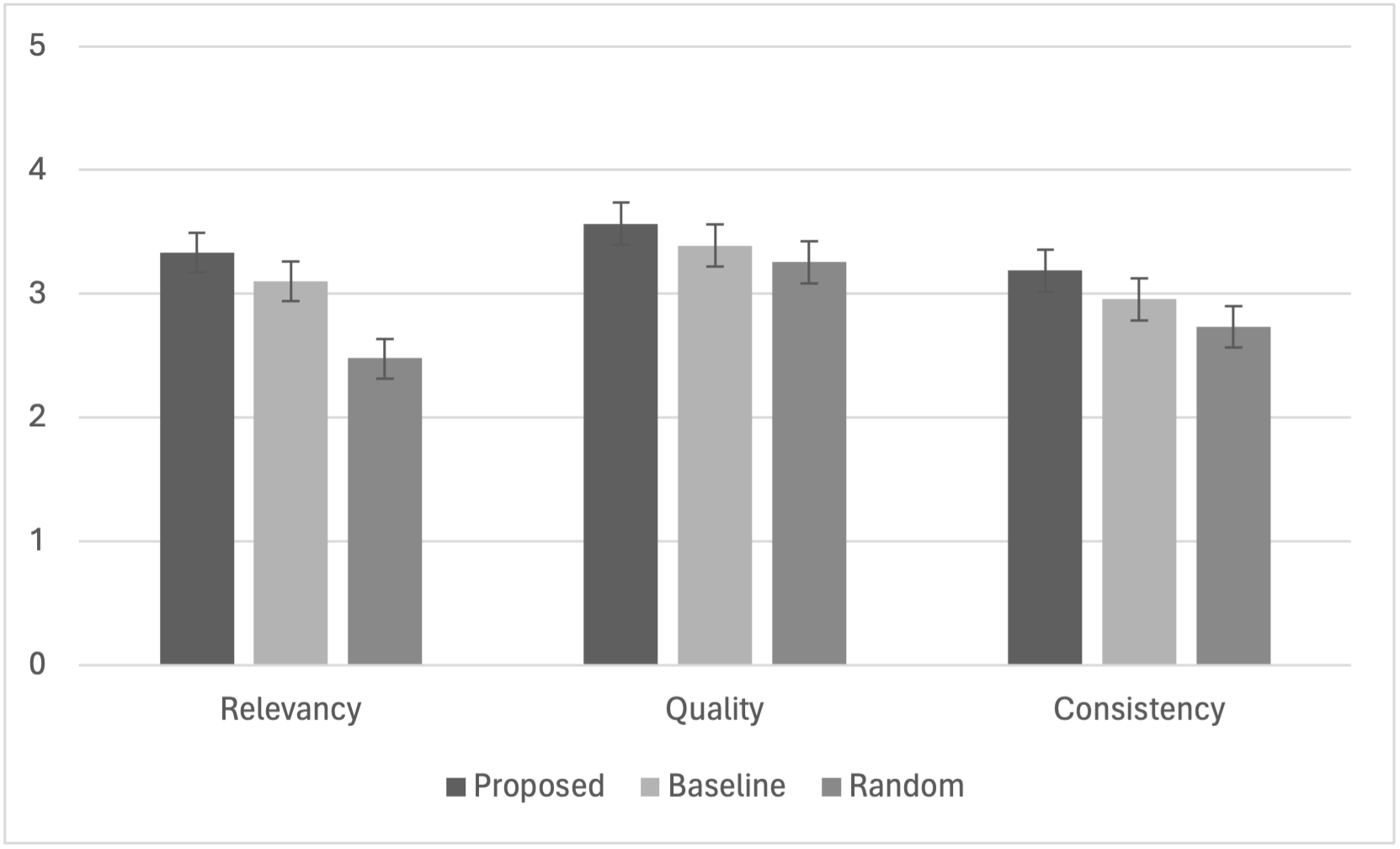}
    \caption{Final mean scores for each of the metric using within-subject ANOVA on the between-scenes study}
    \label{fig:betweenscenesresult}
\end{figure}

\textit{\textbf{Comparison of Within-Scene and Between-Scenes Music Generation.}} The initial survey focused on music generation \textit{within} a single manga scene. To evaluate how M2M-Gen performs when transitioning \textit{between} scenes, where the narrative context may shift entirely, we conducted a follow-up survey. In this study, the M2M-Gen pipeline takes context from two different scenes (but consecutive pages) of six different manga books. There were a total of 15 participants and the survey was designed to be identical to the first. The results for between-scenes study are illustrated in Fig. \ref{fig:betweenscenesresult}. 

This comparison evaluates M2M-Gen’s effectiveness in maintaining musical coherence during scene transitions and its adaptability to narrative shifts. Therefore, \textit{consistency} of the generated music is of particular importance. We find that M2M-Gen generates music with statistically significant consistency $(p < 0.001)$ across changing scenes, effectively capturing contextual nuances and ensuring more coherent and satisfying musical transitions. This is further evidenced by the higher preference scores for M2M-Gen in both within-scene and between-scenes conditions. These findings not only validate our proposed method but also highlight its superiority in delivering consistent and contextually appropriate music, thereby enhancing reading experience.

\subsection{Objective Evaluation} 
Due to the lack of a complete end-to-end dataset, our objective evaluation is limited to assessing the individual components of the pipeline. This subsection focuses on evaluating the performance of CLIP, used for emotion recognition, which is the sole fine-tuned model used in M2M-Gen. The results of fine-tuning CLIP are shown in Table \ref{tab:clipperformance}. Fine-tuning CLIP significantly improves performance compared to its zero-shot counterpart, highlighting its necessity for the task. Additionally, we compare our results with those of Theodose et al. \cite{theodose2023kangaiset}, who used ResNet to achieve comparable results.

\begin{table}[]
\centering
\caption{Comparison of baseline \cite{theodose2023kangaiset} with zero-shot and fine-tuned CLIP models, showing comparable performance to the best results reported by Theodose et al. \cite{theodose2023kangaiset}.}
\begin{tabular}{llll}
          & \textbf{Micro}  & \textbf{Macro}  & \textbf{Weighted} \\ \hline
Baseline  & \textbf{70.9\%} & 40.9\%          & 69.7\%            \\ \hline
Zero-shot & 10.3\%          & 8.9\%           & 10.1\%            \\ \hline
Fine-tuned & 68.7\%          & \textbf{45.8\%} & \textbf{70.2\%}  
\end{tabular}

\label{tab:clipperformance}
\end{table}

\section{Conclusion}
This paper presents a system that can effectively create background, non-diegetic music for Japanese manga. Through extensive experimentation, we have developed a pipeline that extracts information from a manga book and translates it into low-level music conditions for each page, maintaining the coherency and relevancy of the generated music across pages and scenes. By converting different modalities to text before generating music, we establish an abstraction layer that reveals how various elements affect the music’s mood and style. The pipeline also integrates state-of-the-art publicly available models, which reduces training requirements in the absence of a dataset. Incorporating scene segmentation, longer context, and prompt engineering, we are able to create a novel reading experience for manga readers by adding music as an additional stimulus. Our proposed model achieves significantly higher human satisfaction across the ratings of relevancy, quality and consistency compared to the random lower bound. 

Although we see this study as a significant advancement in multi-modal background music generation for manga, offering a foundation for future research and artistic exploration, it does have some limitations. For example, a lack of objective metrics to compare the performance of the proposed model against the baselines using an end-to-end dataset. We also recognise that our pipeline can experience other limitations, such as error propagation to subsequent stages --- e.g. incorrect emotion recognition may weaken the relevancy of the music. Moreover, AI-generated music is still at risk of resembling copyrighted music, which can raise concerns about the ethical and legal use of such models.

Since there is little work with Manga-Music datasets, we find data curation to be a large portion of the future work. Moreover, we plan to improve and evaluate the modules in the pipeline to achieve better narrative consistency. Finally, the world of manga utilises text creatively as visual art. This is most importantly exemplified with onomatopoeia, a style of connecting visual cues with phonetic phrases. We plan to integrate onomatopoeia to enhance the storytelling nature of music as an accompaniment. 

\section*{Acknowledgements}
This work was made possible by the Aizawa Yamakata Matsui Lab at the University of Tokyo, who curated the Manga109 dataset.

\bibliographystyle{IEEEtran}
\bibliography{ref}
\balance

\end{document}